\documentclass[useAMS]{mn2e}

\usepackage{graphicx}
\usepackage{times}
\newcommand{\Teff}{\ensuremath{T_{\rm eff}}}                            
\newcommand{\logg}{\ensuremath{\log g}}                                 
\newcommand{\Msun}{\ensuremath{\,{\rm M}_\odot}}                        
\newcommand{\Rsun}{\ensuremath{\,{\rm R}_\odot}}                        
\newcommand{\ion}[2]{{#1}\,{\sc {\small{#2}}}}                          
\newcommand{\Apx}{\,\AA\,px$^{-1}$}                                     
\newcommand{\kms}{\,km\,s$^{-1}$}                                       
\newcommand{\micro}{\ensuremath{v_{\rm turb}}}                          

\title[Chemical evolution of high-mass stars. I.]
{Chemical evolution of high-mass stars in close binaries. I. The eclipsing binary V453\,Cygni}

\author[K. Pavlovski and  J. Southworth]
{K. Pavlovski$^1$ and J. Southworth$^2$  \\
$^1$\,Department of Physics, University of Zagreb, Bijeni\v{c}ka cesta 32, 10000 Zagreb, Croatia \\
$^2$\,Department of Physics, University of Warwick, Coventry CV7 4AL, UK}

\date{}

\pagerange{\pageref{firstpage}--\pageref{lastpage}} \pubyear{2008}

\begin{document} \maketitle \label{firstpage}

\begin{abstract}
The eclipsing and double-lined spectroscopic binary system V453\,Cygni consists of two early
B-type stars, one of which is nearing the terminal age main sequence and one which is roughly
halfway through its main sequence lifetime. Accurate measurements of the masses and radii of
the two stars are available, which makes a detailed abundance analysis both more interesting
and more precise than for isolated stars. We have reconstructed the spectra of the individual
components of V453\,Cyg from the observed composite spectra using the technique of spectral
 disentangling. From these disentangled spectra we have obtained improved effective temperature
measurements of $27\,900 \pm 400$\,K and $26\,200 \pm 500$\,K, for the primary and secondary
stars respectively, by fitting non-LTE theoretical line profiles to the hydrogen Balmer lines.
Armed with these high-precision effective temperatures and the accurately known surface
gravities of the stars we have obtained the abundances of helium and metallic elements.
A detailed abundance analysis of the primary star shows a normal (solar) helium abundance
if the microturbulence velocity derived from metallic lines is used. The elemental abundances
show no indication that CNO-processed material is present in the photosphere of this high-mass
terminal age main sequence star. The elemental abundances of the secondary star were derived by
differential study against a template spectrum of a star with similar characteristics. Both the
primary and secondary components display elemental abundances which are in the ranges observed
in the Galactic OB stars.
\end{abstract}



\section{Introduction}

High-mass stars are major producers of ultraviolet light and byproducts of thermonuclear fusion,
so are important objects in the study of star formation, and galactic chemical and kinetic
evolution (e.g.\ Samland 1998). The large luminosities of high-mass stars means that they are
important distance indicators in both a galactic and cosmological context (e.g.\ Guinan et al.\
1998; Ribas et al.\ 2005; Southworth et al.\ 2005, 2007).

In the last decade new theoretical models have been constructed which incorporate a number
of physical processes important to the structure and evolution of high-mass stars, including
convective core overshooting, semiconvection, rotational mixing and mass loss. Overshooting
and rotation can both have a large effect on the predicted lifetimes and luminosities of
high-mass stars (c.f.\ review by Maeder \& Meynet 2000). However, empirical constraints on
these processes remain hard to come by, despite a steady improvement in observational
techniques and capabilities (see Hilditch 2004).

Detached eclipsing binaries (dEBs) are vital objects for obtaining observational constraints
on the structure and evolution of high-mass stars, since they are the primary source of
directly measured stellar properties (Andersen 1991). Unfortunately, accurate (2\% or better)
physical properties are available for only eight high-mass dEBs\footnote{An up-to-date
compilation of the properties of well-studied dEBs is available at
{\tt http://www.astro.keele.ac.uk/$\sim$jkt}}, and most of these objects have
no observational constraints on their chemical composition. Chemical abundances
are difficult to determine from the spectra of high-mass dEBs for several reasons.
Firstly, they tend to display only a small number of spectral lines. Secondly, the
often high rotational velocities of these objects means their spectral lines are wide
and shallow so high signal-to-noise ratio (S/N) spectra are needed to obtain useful
results. Thirdly, in dEBs the spectral lines from one component interfere with the
line profiles of the other component (`line blending').

In a seminal work, Simon \& Sturm (1994) introduced the technique of {\em spectral
disentangling} ({\sc spd}), by which {\em individual} spectra of the component stars
of a double-lined spectroscopic binary system can be deduced from a set of composite
spectra observed over a range of orbital phases. {\sc spd} can be used to measure
spectroscopic orbits which are not affected by the line blending which afflicts other
methods, such as cross-correlation (see Southworth \& Clausen 2007). The resulting
disentangled spectra also have a much higher S/N than the original observations,
making them very useful for chemical abundance analyses. As a bonus, the strong
degeneracy between effective temperature (\Teff) and surface gravity (\logg) is
not a problem for dEBs because surface gravities can be measured to high accuracy
(0.01\,dex or better) for both components. A detailed study of these possibilities
is given by Pavlovski \& Hensberge (2005; hereafter PH05).

\subsection{The eclipsing binary system V453\,Cyg}   \label{sec:intro:v453}

V453\,Cyg belongs to rather sparse group of high-mass close binary systems for which accurate
physical properties have been measured for both components. In the most recent study of
V453\,Cyg, Southworth, Maxted \& Smalley (2004b; hereafter SMS04) analysed new spectroscopic
observations and previously published light curves, obtaining masses of $M_{\rm A} = 14.36
\pm 0.20$\Msun\ and $M_{\rm B} = 11.11 \pm 0.13$\Msun, and radii of $R_{\rm A} = 8.55 \pm
0.06$\Rsun\ and $R_{\rm B} = 5.49 \pm 0.06$\Rsun. The high precision of the radius measurements
was helped by the geometry of the V453\,Cyg system, which exhibits deep and total eclipses.
SMS04 derived effective temperatures of ${\Teff}_{\rm A} = 26\,500 \pm 800$\,K and ${\Teff}_{\rm B}
= 25\,300 \pm 600$\,K from the equivalent widths of several \ion{He}{I} and \ion{He}{II} lines,
using an analysis based on local thermodynamic equilibrium (LTE) synthetic spectra but with
a correction for non-LTE effects.

V453\,Cyg is known to display the phenomenon of apsidal motion, which can be used as a probe of the
internal structure of the components of an eccentric binary system (e.g.\ Claret \& Gim\'enez 1993).
SMS04 found an apsidal period of $U = 66.4 \pm 1.8$\,yr, which gives the structural
parameter\footnote{Note that the $\log k_2$ value in the abstract of SMS04 is incorrect: the correct
value is given in SMS04 (their Section\,7.2) and here.} $\log k_2 = -2.254 \pm 0.024$. The masses,
radii, \Teff s and $\log k_2$ of V453\,Cyg match theoretical predictions for an age of $10.0 \pm
0.2$\,Myr and an approximately solar metal abundance. The primary component (the hotter and more
massive star) is reaching the terminal age of its main sequence lifetime (TAMS) whilst the secondary
star is roughly halfway through its MS life. The significant difference between the properties
of the two stars means that V453\,Cyg is a very useful system for testing evolutionary models.

In this work we analyse the spectroscopy of V453\,Cyg obtained by SMS04 and by Simon \& Sturm
(1994) in order to obtain new \Teff s and accurate chemical abundances for the stars. V453\,Cyg
was targeted by SMS04 as part of a program to study dEBs which are members of open clusters
(see Southworth, Maxted \& Smalley 2004a; Southworth et al.\ 2004c). It is a probable member of
NGC\,6871, so our results can be compared to published analyses of single stars which are also
cluster members.


\section{Spectroscopic data}          \label{sec:obs}

\begin{table} \centering
\caption{\label{tab:obs:caha} Observing log for the Calar Alto spectra
of V453\,Cyg. $\lambda_{\rm c}$ is the central wavelength and the
orbital cycles are calculated from HJD 2\,453\,940.0998.}
\begin{tabular}{cccccc} \hline
Set & ID & HJD & S/N  & $\lambda_{\rm c}$ (\AA) & Cycle \\
\hline
A   & 21x   & 48813.39567  & 150  &  4010  & 2435.4036  \\
A   & 43x   & 48814.63754  & 150  &  4010  & 2435.7229  \\
A   & 51x   & 48816.38687  & 160  &  4010  & 2436.1726  \\
A   & 61x   & 48817.38987  & 170  &  4010  & 2436.4304  \\
A   & 71x   & 48818.37820  & 190  &  4010  & 2436.6845  \\
A   & 81x   & 48819.38231  & 190  &  4010  & 2436.9427  \\
A   & 32    & 49201.45130  & 140  &  4010  & 2535.1653  \\
A   & 41    & 49202.35881  & 170  &  4010  & 2535.3986  \\
A   & 51    & 49203.46754  & 190  &  4010  & 2535.6837  \\
\hline
B   & 22x   & 48813.47704  & 180  &  4290  & 2435.4245  \\
B   & 42x   & 48815.46905  & 170  &  4290  & 2435.9366  \\
B   & 62x   & 48816.52872  & 220  &  4290  & 2436.2091  \\
B   & 72x   & 48817.64888  & 200  &  4290  & 2436.4970  \\
B   & 85x   & 48818.60773  & 190  &  4290  & 2436.7435  \\
B   & 82x   & 48819.47156  & 190  &  4290  & 2436.9656  \\
\hline
C   & 23x   & 48812.57408  & 180  &  4520  & 2435.1924  \\
C   & 32x   & 48814.48183  & 180  &  4520  & 2435.6828  \\
C   & 41x   & 48815.40237  & 140  &  4520  & 2435.9195  \\
C   & 52x   & 48816.42184  & 190  &  4520  & 2436.1816  \\
C   & 63x   & 48816.63435  & 210  &  4520  & 2436.2362  \\
C   & 83x   & 48818.59129  & 200  &  4520  & 2436.7393  \\
\hline
D   & 11x   & 48812.48692  & 200  &  4750  & 2435.1700  \\
D   & 11    & 49199.49495  & 200  &  4750  & 2534.6624  \\
D   & 12    & 49199.60984  & 190  &  4750  & 2534.6919  \\
D   & 13    & 49199.65734  & 150  &  4750  & 2534.7041  \\
D   & 21    & 49200.38908  & 170  &  4750  & 2534.8923  \\
D   & 31    & 49201.41599  & 180  &  4750  & 2535.1562  \\
D   & 33    & 49201.66650  & 150  &  4750  & 2535.2207  \\
D   & 42    & 49202.38781  & 170  &  4750  & 2535.4061  \\
\hline \end{tabular} \end{table}

\begin{table} \centering
\caption{\label{tab:obs:int} Observing log of La Palma spectra of
V453\,Cyg. Columns are the same as for Table\,\ref{tab:obs:caha}.}
\begin{tabular}{cccccc} \hline
Set & ID & HJD & S/N & $\lambda_{\rm c}$ (\AA) & Cycle \\
\hline
E  &  323747  &  52562.50016  &  160  &  4360  &  3399.2270    \\
E  &  323867  &  52563.31850  &  110  &  4360  &  3399.4374    \\
E  &  324029  &  52564.31988  &  160  &  4360  &  3399.6948    \\
E  &  324255  &  52565.33945  &  180  &  4360  &  3399.9569    \\
E  &  324455  &  52566.29608  &  140  &  4360  &  3400.2029    \\
E  &  324462  &  52566.31116  &  140  &  4360  &  3400.2068    \\
E  &  324505  &  52566.43565  &  180  &  4360  &  3400.2388    \\
E  &  324581  &  52568.34709  &  160  &  4360  &  3400.7302    \\
E  &  324757  &  52569.31004  &  100  &  4360  &  3400.9777    \\
E  &  324774  &  52569.34031  &  130  &  4360  &  3400.9855    \\
\hline
F  &  323085  &  52559.33775  &  210  &  4580  &   3398.4140   \\
F  &  323086  &  52559.34009  &  210  &  4580  &   3398.4146   \\
F  &  323285  &  52560.31872  &  150  &  4580  &   3398.6662   \\
F  &  323286  &  52560.32104  &  150  &  4580  &   3398.6668   \\
F  &  323287  &  52560.32335  &  150  &  4580  &   3398.6674   \\
F  &  323288  &  52560.32567  &  150  &  4580  &   3398.6680   \\
F  &  323289  &  52560.32799  &  150  &  4580  &   3398.6686   \\
F  &  323317  &  52560.40911  &  130  &  4580  &   3398.6894   \\
F  &  323318  &  52560.41142  &  130  &  4580  &   3398.6900   \\
F  &  323319  &  52560.41375  &  130  &  4580  &   3398.6906   \\
F  &  323320  &  52560.41606  &  130  &  4580  &   3398.6912   \\
F  &  323321  &  52560.41837  &  130  &  4580  &   3398.6918   \\
F  &  323479  &  52561.30395  &  140  &  4580  &   3398.9195   \\
F  &  323480  &  52561.30627  &  150  &  4580  &   3398.9201   \\
F  &  323481  &  52561.30858  &  150  &  4580  &   3398.9207   \\
F  &  323482  &  52561.31091  &  140  &  4580  &   3398.9213   \\
F  &  323483  &  52561.31322  &  150  &  4580  &   3398.9219   \\
F  &  323744  &  52562.49486  &  160  &  4580  &   3399.2257   \\
F  &  323864  &  52563.31216  &  110  &  4580  &   3399.4358   \\
F  &  324026  &  52564.31111  &  170  &  4580  &   3399.6926   \\
F  &  324247  &  52565.32035  &  180  &  4580  &   3399.9520   \\
F  &  324584  &  52568.35585  &  170  &  4580  &   3400.7324   \\
F  &  325129  &  52570.35446  &  170  &  4580  &   3401.2462   \\
F  &  325348  &  52571.30720  &  180  &  4580  &   3401.4911   \\
\hline \end{tabular} \end{table}

\subsection{Calar Alto spectra}

A total of 29 spectra of V453\,Cyg were obtained by Simon \& Sturm (1994) during two observing
runs in 1990 and 1992 with the 2.2\,m telescope at the German-Spanish Astronomical Centre
at Calar Alto, Spain (Table\,\ref{tab:obs:caha}). Four different spectral intervals were
observed with the coud\'{e} spectrograph, each covering 250\,\AA. The intervals are centred
on the hydrogen lines H$\delta$ ($\lambda$4050), H$\gamma$ ($\lambda$4430), H$\beta$
($\lambda$4850), and on a 4533--4730\,\AA\ region containing metallic lines, \ion{He}{II}
$\lambda$4686 and \ion{He}{I} $\lambda$4712. The reciprocal dispersion is 0.13\Apx\ for
the 1990 spectra and 0.20\Apx\ for the 1992 data. The S/N ratios are 140--220; further
details are given by Simon et al.\ (1994).

\subsection{La Palma spectra}

A set of 43 spectra of V453\,Cyg were secured in 2001 October with the 2.5\,m Isaac Newton
Telescope and Intermediate Dispersion Spectrograph at La Palma, Spain (Table\,\ref{tab:obs:int}).
These are split between two spectral intervals of length 260\,\AA, reciprocal dispersion 0.11\Apx\
and resolution 0.20\,\AA. Most spectra are centred on 4360\,\AA\ and the remainder on 4580\,\AA:
both datasets cover the \ion{He}{I} $\lambda$4471 and \ion{Mg}{II} $\lambda$4481 lines, which
are good lines for radial velocity (RV) determination over a wide range of spectral types
(Andersen 1975). A few additional spectra were obtained with central wavelengths of 4340\,\AA\
and 4840\,\AA; they are not used here. The S/N of the La Palma spectra is 100--450 per pixel,
with most close to 150. We have rejected a few spectra with a S/N below 100.


\section{Method}

Our aim is to perform a detailed test of the predictions of stellar evolution models of
high-mass stars using strict empirical constraints. This aim can only be reached by studying
dEBs, as these are the only objects where we can measure masses and radii of high-mass stars
to accuracies of 1\%. In this context the {\sc spd} technique has an important role both in
mass and abundance determination. RV and mass measurements of OB-type dEBs are difficult because
their spectra generally contain only a few lines, which have strong rotational
broadening and so suffer line blending.

Using {\sc spd} and a set of spectra taken at a range of orbital phases,
it is possible to determine the individual component spectra. One can also simultaneously solve for
the RVs for each component and spectrum, or directly for the orbital parameters of the system.
Contrary to the cross-correlation method (Simkin 1974; Tonry \& Davis 1979) no template spectrum
is needed since the individual component spectra are themselves used as templates, so a significant
source of bias is bypassed (c.f.\ Hensberge \& Pavlovski 2007).

Several studies have put extensive effort into reconciling RVs measured by several techniques
to orbital parameters derived using {\sc spd}. Orbital parameters from RVs determined by
multi-component Gaussian fitting to complex line profiles have been found to be inferior to
those derived by {\sc spd} (Simon \& Sturm 1994b; Simon, Sturm \& Fiedler 1994; Hensberge et
al.\ 2000). {\sc spd} also performs better than standard cross-correlation (Holmgren et al.\ 1997).

Southworth \& Clausen (2007) critiqued these methods and the two-dimensional cross-correlation
technique {\sc todcor} (Zucker \& Mazeh 1994) in an analysis of the dEB DW\,Carinae. The main
problem in RV determination was found to be line blending. {\sc spd} gave the most
reliable RVs, and fitting a double-Gaussian also gave acceptable results. Cross-correlation
techniques could not properly account for line blending and gave results which were
systematically biased towards smaller stellar masses. These conclusions were obtained
in a case when the two component stars had very similar properties and individual spectral
lines were studied, so some of the benefits of cross-correlation could not be accessed.

\begin{figure*} \includegraphics[width=160mm]{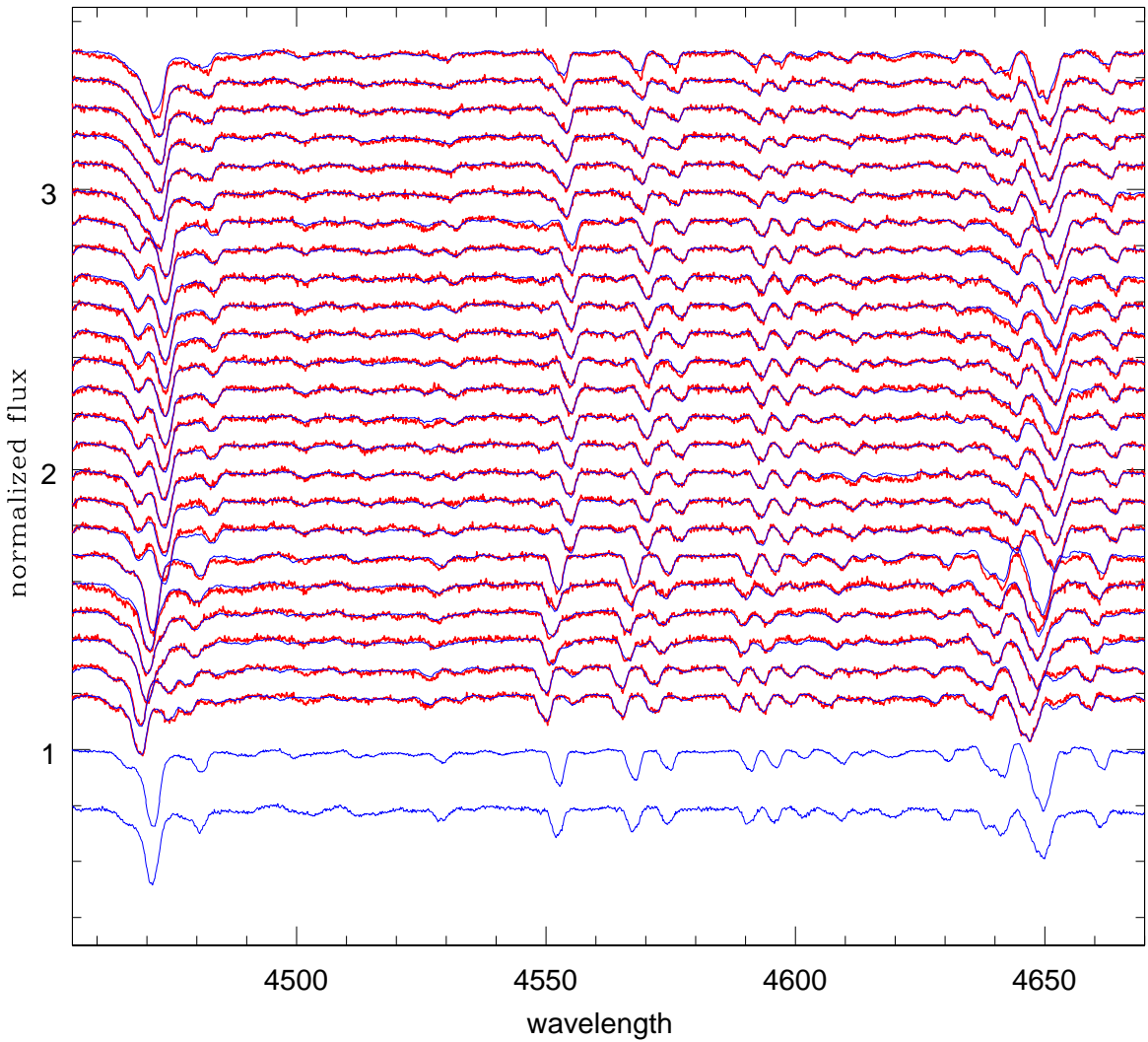} \\
\caption{\label{fig:example} Example region of the observed composite spectra
(red lines) from La Palma subset\,F. The two individual disentangled spectra
are shown at the bottom of the plot (blue lines) with their correct continuum
levels. The disentangled spectra have then been adjusted with the appropriate
Doppler shifts and relative intensities to reproduce the observed spectra, and
are overlaid on them using blue lines. All spectra have been normalised to the
continuum, and the observed spectra have each been shifted by $+0.1$ from their
neighbour for clarity. The spectrum taken during secondary eclipse is the sixth
from the bottom among the observed spectra.} \end{figure*}


\section{Spectroscopic orbits through spectral disentangling}           \label{sec:orbits}

\begin{table} \centering
\caption{\label{tab:orbit} Parameters of the spectroscopic orbits derived in
this work and by SMS04. SMS04 fixed the orbital period ($P = 3.889825$\,d),
time of periastron passage ($T_{\rm peri} = 2\,439\,340.6765$) and eccentricity
($e$) to values found in their apsidal motion analysis. In the present work we
have optimised $e$, the longitude of periastron ($\omega$) and the amplitudes
of RV variation ($K_{\rm A}$ and $K_{\rm B}$). }
\begin{tabular}{lccc} \hline
Parameter               & SMS04               & Subset C            & Subset F           \\
\hline
$e$                     & 0.022 (fixed)       & $0.029 \pm 0.016$   & $0.024 \pm 0.003$  \\
$\omega$ ($^\circ$)     & 140.1 (fixed)       & $153.4 \pm 13.7$    & $166.7 \pm 8.6$    \\
$K_{\rm A}$ (\kms)      & $173.7 \pm 0.8$     & $174.1 \pm 0.9$     & $172.5 \pm 0.2$    \\
$K_{\rm B}$ (\kms)      & $224.6 \pm 2.0$     & $216.2 \pm 1.7$     & $221.5 \pm 0.5$    \\
$q$                     & $0.773 \pm 0.008$   & $0.805 \pm 0.008$   & $0.779 \pm 0.003$  \\
\hline \end{tabular} \end{table}

Two spectroscopic datasets are available for analysis which were obtained about 10 years apart
(Section\,\ref{sec:obs}). As V453\,Cyg exhibits fast apsidal motion (see Section\,\ref{sec:intro:v453})
these cannot be analysed together. The datasets are split over a number of spectral windows:
subset\,F (La Palma) is the most extensive and includes 24 spectra centred on 4580\,\AA. This
subset was used for determination of the orbital parameters. The phase distribution of the
spectra is reasonable but suffers slightly from the orbital period of V453\,Cyg (3.89\,d)
being close to an integer number of days: phases $\phi = 0.0$--$0.20$ could not be covered.
However, an observation was deliberately obtained during secondary eclipse, where only the
primary star is visible. This spectrum was actually used as the template in the {\sc todcor}
analysis performed by SMS04 as it is a good match to both stars. Disentangled spectra formally
contain no information about the continuum level for each star (PH05), but in the case of
V453\,Cyg this secondary-eclipse spectrum can be used to reliably determine how much light
each star contributes.

We have performed a spectral disentangling analysis in velocity space (see Hadrava 1995) using
the {\sc fdbinary}\footnote{\tt http://sail.zpf.fer.hr/fdbinary/} code (Iliji\'{c} et al.\ 2004).
In Fourier {\sc spd} individual pixel flux uncertainties cannot be used but individual
spectra can be assigned a weight based on their quality. We weighted the input spectra according
to their S/N and adopted as our initial orbit the final solution given by SMS04. We fixed the
orbital period at 3.889825\,d (SMS04).

{\sc spd} was performed over the spectral region 4540--4690\,\AA. The strong lines of 
\ion{He}{II} $\lambda$4471 and \ion{Mg}{II} $\lambda$4481 were excluded in the determination 
of the orbital parameters, as in our experience better results are obtained for larger numbers 
of weak lines than for a few strong lines. The resulting parameters are given in Table\,\ref{tab:orbit}.
 The errors are determined by the jackknife method (e.g., Yang \& Robinson 1986), which seems to be 
an appropriate approach for error calculations in {\sc spd}. Table\,\ref{tab:orbit} also contains 
the final orbit of SMS04, which is in good agreement with the results found here. Note that, in 
their analysis, SMS04 used {\sc todcor} to measure RVs so were only able to use spectra where the
 two stars had very different RVs. They thus fixed the eccentricity, $e$, and periastron longitude, 
$\omega$, to the values found from their apsidal motion analysis. In our solution we were able to use 
all the available 4580\,\AA\ spectra (subset\,F), including that obtained during secondary eclipse 
(phase 0.50). The orbital parameters are therefore more reliable, particularly $e$ and $\omega$. 
An example of {\sc spd} of the observed composite spectra and resulting disentangled spectra are 
shown in Fig.\,\ref{fig:example}.

For the Calar Alto spectra, subset\,C was chosen to be the master dataset for determining the
 orbital parameters. Since the individual Calar Alto datasets contain fewer spectra than the La Palma
 ones, we fixed the eccentricity to the value obtained from subset\,F ($e = 0.023$). Other parameters 
were adjusted in {\sc spd}. Other subsets (A, B, and D) were disentangled keeping the orbital parameters
 fixed to those obtained for subset\,C. The La Palma spectra were disentangled using the orbital 
parameters obtained for subset\,F.

A feature of {\sc spd} is that the disentangled spectra are effectively co-added from the original 
observations and so have a higher S/N. The total S/N in a disentangled spectrum of component $n$
 is $$ {\rm S_{dis}} \sim f_n {\rm S_{obs}} \sqrt{N_{\rm obs}} $$ where $f_n$ is the fractional 
light contribution for component $n$, and ${\rm S_{obs}}$ and $N_{\rm obs}$ are the S/N and number 
of input spectra. Both sets of the La Palma spectra cover the 4455--4505\,\AA\ region (which includes
 \ion{He}{I} $\lambda$4471 and \ion{Mg}{II} $\lambda$4481), so the total number of observations of 
this spectral window is 45. We have disentangled this region separately, yielding individual component
 spectra with S/N of about 700 for the primary and 250 for the secondary ($f_{\rm A} = 0.75$ and 
$f_{\rm B} = 0.25$). For comparison, disentangling the eight spectra in subset\,C (Calar Alto) gives
 individual spectra with S/N of about 380 and 130. The enhanced S/N in the disentangled spectra of
 the components relative to the observed spectra can be seen in Fig.\,\ref{fig:example}. The two 
disentangled spectra have different S/N because the secondary star is fainter than the primary.


\section{Spectral analysis of both components}

The primary goal of this study is to measure the photospheric elemental abundances of the
components of V453\,Cyg, which requires knowledge of their \Teff s and \logg s. Whilst
disentangled spectra can be analysed in the same ways as single-star spectra, a potential
drawback is that the continuum level for each star is not specified by {\sc spd},
short of ensuring individual spectral lines do not dip below zero flux. For V453\,Cyg this
is not a problem because spectra have been obtained during eclipses (spectrum 324777 in
subset\,F and spectrum 11x in subset\,D). Also, the solution of the Cohen (1974) light
curves of V453\,Cyg (SMS04) allows us to assign a light factor to the two stars for each
observed spectrum. We have renormalised the disentangled spectra using the method and formulae
 presented by PH05. The mid-eclipse spectra are themselves suitable for the following analysis,
but we have used the renormalised disentangled spectra due to their much higher S/N.

Abundance analyses of single stars can suffer from a strong correlation between the \Teff\
and \logg\ measurements from stellar spectra. In the case of dEBs this can be
avoided because the \logg\ of each star can be measured accurately. The values for V453\,Cyg
are ${\logg}_{\rm A} = 3.731 \pm 0.012$ and ${\logg}_{\rm B} = 4.005 \pm 0.015$ (SMS04).

\begin{figure*}
\begin{tabular}{ccc}
\includegraphics[width=56mm]{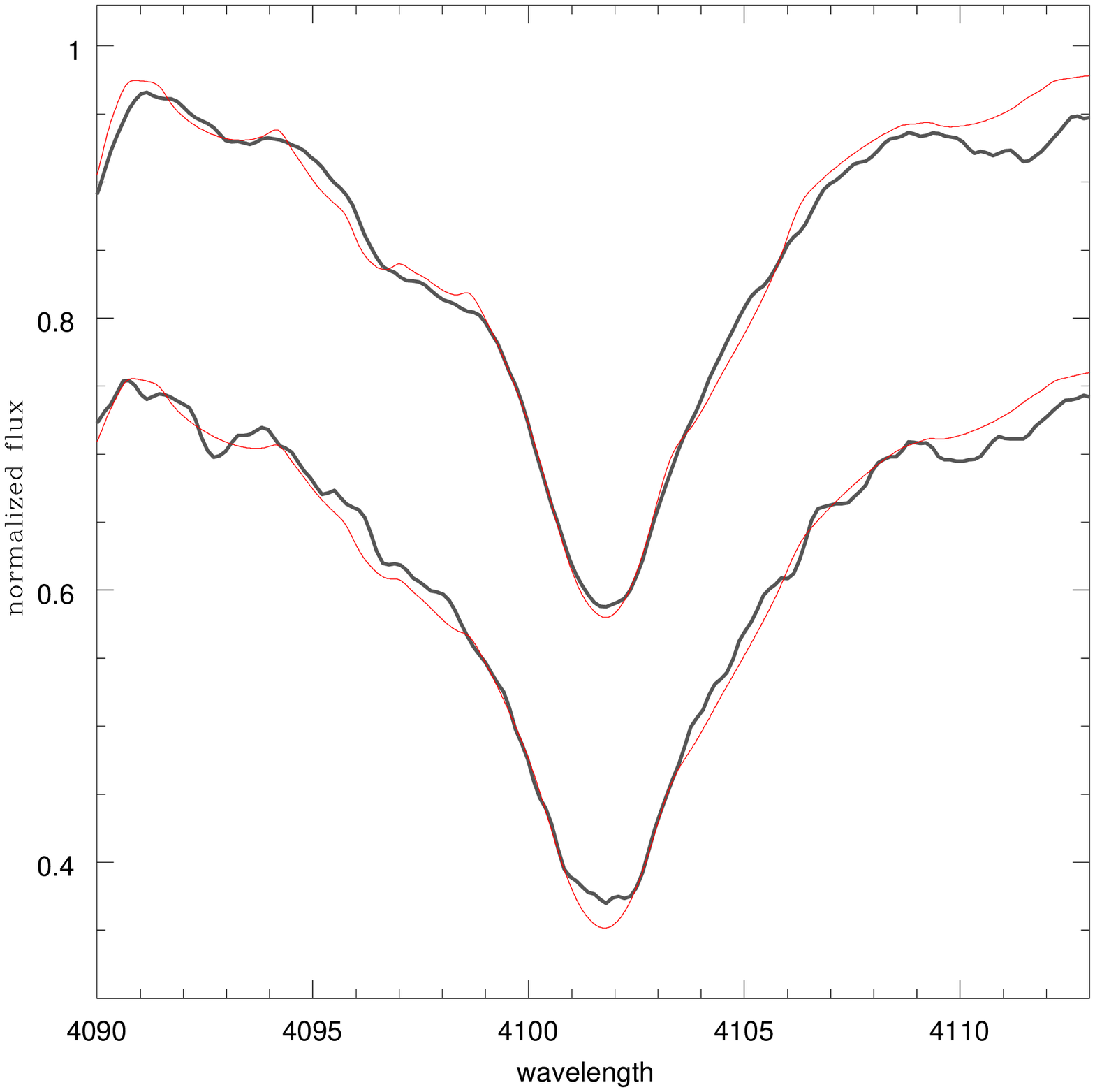} &
\includegraphics[width=56mm]{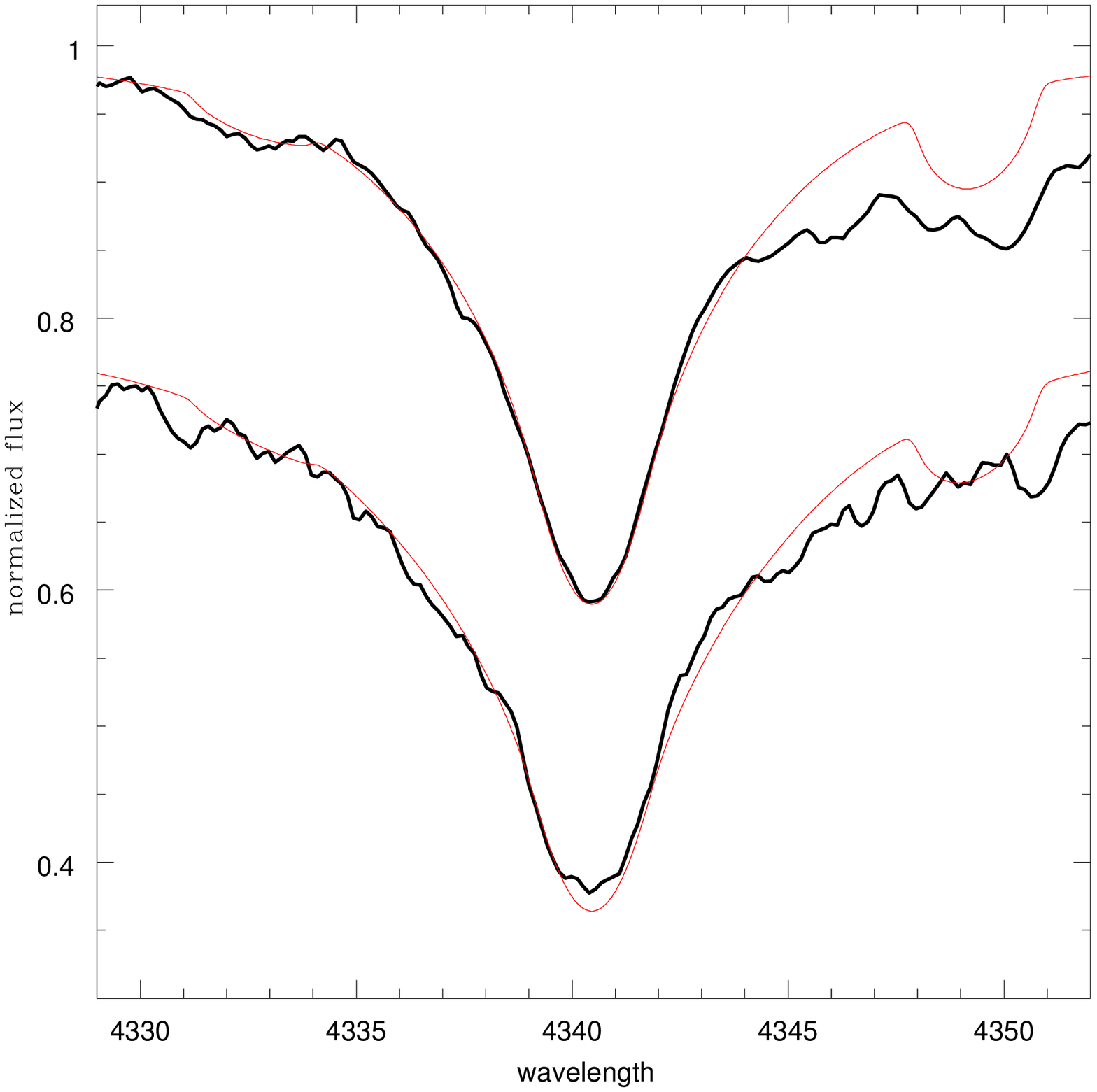} &
\includegraphics[width=56mm]{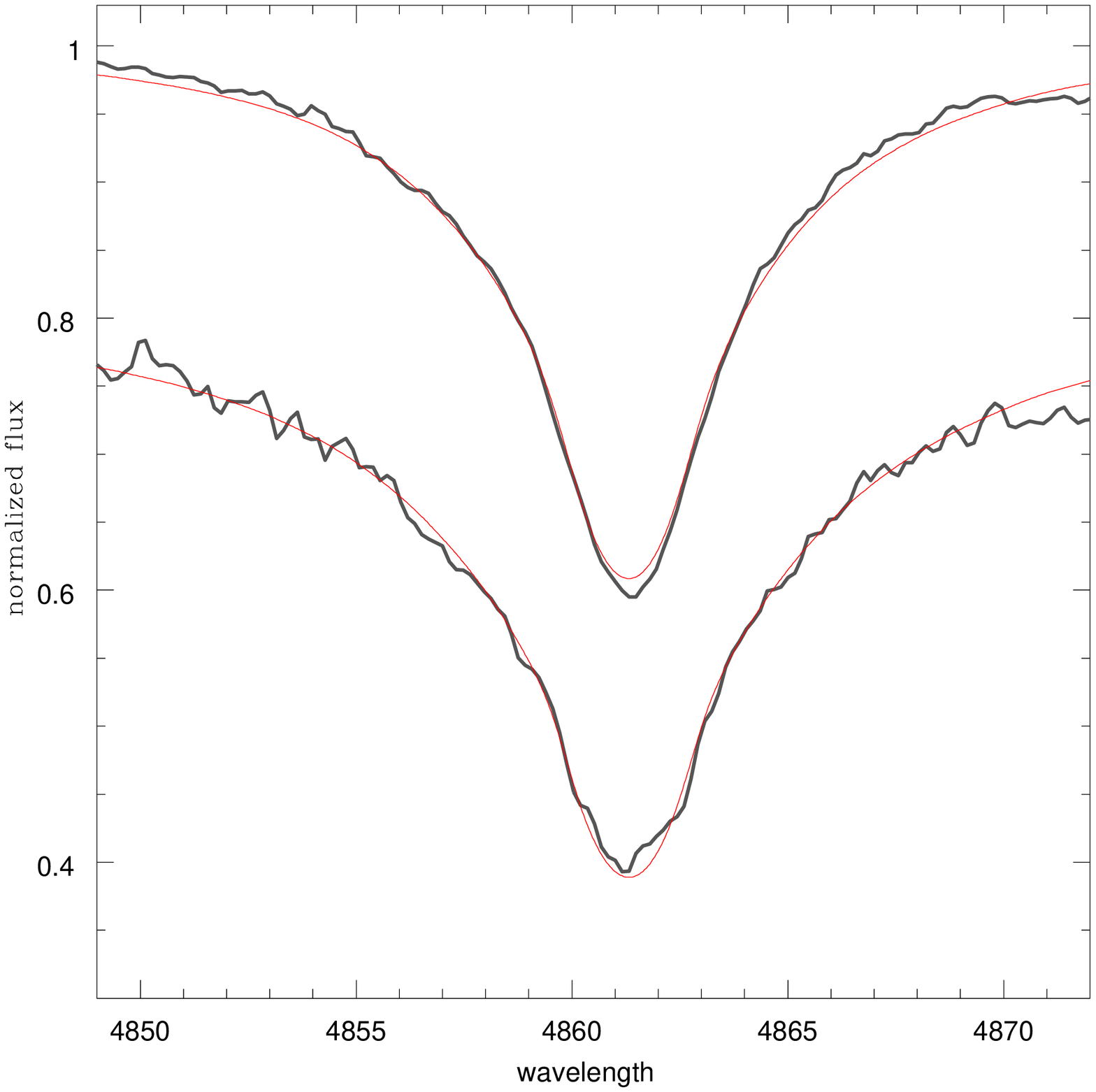} \\
\end{tabular}
\caption{\label{fig:hbgfit} Comparison between the disentangled spectra
(points) and best-fitting theoretical spectra (solid lines) for H$\delta$
(left panel), H$\gamma$ (middle panel) and H$\beta$ (right panel). Spectra
for the secondary star have been offset by $-0.2$ for clarity.}
\end{figure*}

\subsection{Effective temperature determination}                              \label{sec:teff}

SMS04 used the equivalent widths of helium lines and LTE theoretical spectra (with a correction
for non-LTE effects) to derive ${\Teff}_{\rm A} = 26\,500 \pm 800$\,K and ${\Teff}_{\rm B} =
25\,300 \pm 600$\,K, but noted that the helium abundance might be enhanced and that this could
affect the measurements. As the helium abundance is an important part of the current analysis,
we could not use helium lines to measure \Teff s.

A standard way of deriving \Teff\ for a sequence of B stars is from the \ion{Si}{II}/\ion{Si}{III}/\ion{Si}{IV}
 ionisation equilibrium (c.f.\ Becker \& Butler 1990). Some spectral lines of these three ions are
 visible in the primary's disentangled spectra, but its relatively high projected rotational velocity 
(110\kms) means the equivalent width of \ion{Si}{II} $\lambda$\,4128 is difficult to measure to 
sufficient accuracy. Also, some important lines (e.g.\ \ion{Si}{II} $\lambda$5379) are not covered
 by the available spectra. In the secondary's spectrum, \ion{Si}{IV} $\lambda$\,4212 is barely visible.

We have instead used the Balmer lines, which are good indicators of \Teff\ in hot stars if
\logg\ is already known from elsewhere. Our disentangled spectra cover H$\beta$, H$\gamma$ and
H$\delta$. In OB stars the Balmer lines are affected by numerous oxygen, nitrogen and
silicon lines, so we carefully preselected `clean' parts suitable for matching to theoretical
line profiles.  Grids of theoretical spectra were calculated for $\Teff = 24\,000$--$29\,000$\,K
and for the known \logg s of the components of V453\,Cyg. For these calculations we used
{\sc atlas9} model atmospheres (Kurucz 1974) with a microturbulence velocity of
$\micro = 2$\kms\ and solar metal abundances. The codes {\sc detail} (Giddings 1981; Butler 1994)
 and {\sc surface} (Butler 1984) were used for the non-LTE calculations -- {\sc detail}
determines the atomic level populations by jointly solving the radiative transfer and
statistical equilibrium equations, and {\sc surface} computes the emergent fluxes and
line profiles. Model atoms were used for H (Herrero 1987; Husfeld et al.\ 1989),
He (Husfeld et al.\ 1989), \ion{C}{II} (Eber \& Butler 1988), \ion{N}{II} (Becker \& Butler 1989),
 \ion{O}{II} (Becker \& Butler 1988), \ion{Mg}{II} (Przybilla et al.\ 2001), \ion{Al}{III}
(Dufton et al.\ 1986), and \ion{Si}{II}, \ion{Si}{III} and \ion{Si}{IV} (Becker \& Butler 1990).

{\sc fortran} programs have been developed which fit theoretical line profile to hydrogen
lines by $\chi^2$ minimisation in \Teff. More weight was given to the H$\gamma$ profile since
 it was derived using a dataset which contains a spectrum taken during mid-eclipse, and is
covered by both the Calar Alto and the La Palma data sets. Parts of the profiles which are
affected by metallic lines were excluded from the fit. The best fits were found for
${\Teff}_{\rm A} = 27\,800 \pm 400$\,K and ${\Teff}_{\rm B} = 26\,200 \pm 500$\,K, and are
plotted in Fig.\,\ref{fig:hbgfit} for the H$\gamma$ and H$\beta$ lines. These \Teff s are
larger by 1.5$\sigma$ and 1.7$\sigma$ than the values found by SMS04, which can be ascribed
to the importance of non-LTE effects at these temperatures. Note that this does not affect
the light or RV curve analyses presented by SMS04.

While not used in deriving the above \Teff s, the lines of \ion{He}{II} were examined for
the reliability of the results. Our data cover three \ion{He}{II} lines, at at $\lambda$4199,
$\lambda$4541 and $\lambda$4686. A broad and diffuse line at $\lambda$4199 is barely visible
in the primary star's spectrum. The lines at $\lambda$4541 and $\lambda$4685 are consistent
with derived primary \Teff\ but are better fitted with $\Teff \sim 27\,500$\,K (0.75$\sigma$
 lower than found above). The match of the \ion{He}{II} line at $\lambda$4686 in
the secondary's spectrum with theoretical profiles for $\Teff = 26\,200$\,K is excellent.

The projected rotational velocities ($v \sin i$) of the components were derived from the widths
 of the clean spectral features due to \ion{C}{II} $\lambda$4267, \ion{Si}{III} $\lambda$4553,
\ion{Si}{III} $\lambda$4568, \ion{O}{II} $\lambda$4591, \ion{O}{II} $\lambda$4596, and \ion{O}{II}
 $\lambda$4662, avoiding \ion{He}{I} and \ion{Mg}{II} lines (Hensberge et al.\ 2000), using a set
 of the theoretical spectra calculated for different $v \sin i$ values. We find $v_{\rm A}
\sin i = 109 \pm 3$\kms\ and $v_{\rm B} \sin i = 98 \pm 5$\kms. These figures are in excellent
agreement with those found by SMS04. The pseudo-synchronous rotational velocities are 110\kms\
and 71\kms; the primary is therefore pseudo-synchronised but the secondary is rotating more
quickly than expected.

In their abundance analysis of disentangled component spectra of the eclipsing and double-lined
 binary V578\,Mon, a member of the Rosette nebula cluster NGC\,2244, PH05 used a sharp-lined
 spectrum of the cluster member star \#201 as a template spectrum. Vrancken et al.\ (1997)
 derived $\Teff = 26\,500 \pm 1000$ K  and $\logg = 4.3 \pm 0.2$ for this star, so it should
be a good match to V453\,Cyg\,B. A rotationally broadened spectrum of \#201 closely resembles
V453\,Cyg\,B, which lends additional support to the \Teff s derived in this work.

As an independent check we have used Str\"omgren $uvby\beta$ calibrations to obtain the mean
 \Teff\ of the system. The $b-y$, $m_1$ and $c_1$ indices given by Crawford \& Barnes (1974)
and Reimann (1989) are in good agreement with each other, indicating that they are not
affected by the photometric variability of V453\,Cyg. The {\sc uvbybeta} code was used
to apply the calibration of Moon \& Dworetsky (1985) to the Str\"omgren indices. We found
a reddening of $E_{b-y} = 0.33$\,mag, which is in excellent agreement with the value of $E_{B-V}
= 0.46 \pm 0.03$\,mag found by Massey, Johnson \& DeGioia-Eastwood (1995) for NGC\,6871,
and a \Teff\ of 27\,720\,K. This \Teff\ value is a flux-weighted mean for the two stars
because we do not have individual Str\"omgren indices for the two components, and is in
good agreement with the values we find from the Balmer line profiles.

\subsubsection{Implications of the new effective temperature measurements}

Armed with the \Teff\ values determined above, we have calculated the distance to V453\,Cyg
using theoretical bolometric corrections published by Bessell, Castelli \& Plez (1998)
and by Girardi et al.\ (2002). We adopted the spectroscopic orbits of the two stars from
Section\,\ref{sec:orbits} and the photometric parameters from SMS04. The two sources of
bolometric corrections are in almost perfect agreement, and lead to a $V$-band distance of
 $d_V = 1720 \pm 80$\,pc and a $K$-band value of $d_K = 1680 \pm 30$\,pc. The uncertainty
 in the reddening towards V453\,Cyg dominates the uncertainty in $d_V$ but is much less
important for $d_K$. This distance is not in good agreement with the value of $2140 \pm 70$\,pc
found for NGC\,6871 by Massey et al.\ (1995), but is consistent with other values in the literature.

We have compared the properties of V453\,Cyg with several sets of theoretical models,
but cannot find a good match for any of the chemical compositions for which models are available.
Interestingly, the best (i.e.\ least bad) match is obtained with the Claret (1995) models
for a `solar' metal abundance ($Z = 0.02$), a high fractional helium abundance of $Y = 0.38$,
and an age of 7.4\,Myr. In this case the \Teff s and the radius of the primary star are
correctly predicted, but the measured radius of the secondary star is a few sigma smaller
than the model value. A new photometric analysis of V453\,Cyg would help to
investigate this discrepancy further.

\begin{figure*} \includegraphics[width=180mm]{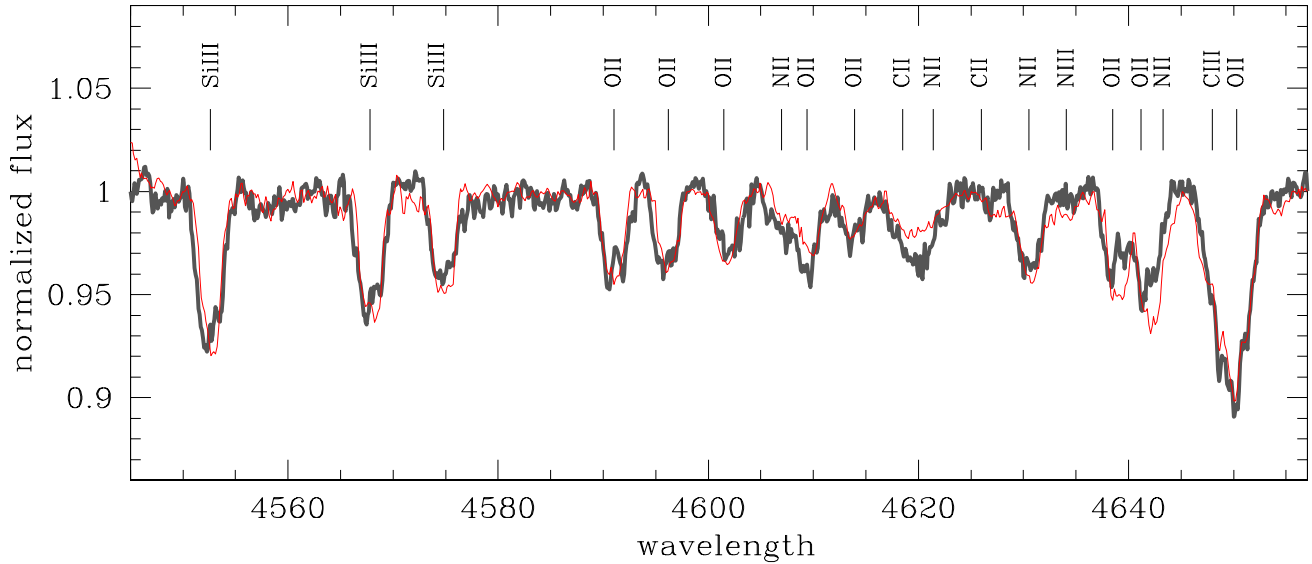}
\caption{\label{fig:v578v453} The spectrum of the secondary component
(solid line) of V578 Mon derived by {\sc spd} (Hensberge et al.\ 2000)
overplotted on the disentangled spectrum of the secondary component of
V453\,Cyg (filled symbols, this work). The stars have almost the same
\Teff, and their \logg s differ by only 0.15\,dex. It is clear that
the stars have very similar abundance patterns.} \end{figure*}

\subsection{Helium abundance} \label{sec:he}

The spectra of early-B stars are rich in strong hydrogen and \ion{He}{I} lines. For the hottest
B stars, \ion{He}{II} lines become visible. The inclusion of non-LTE effects is important when
 determining the helium abundance of such stars, but its importance varies for different lines.
The {\sc detail} and {\sc surface} codes use a model atom calculated including transitions up
to $n=4$ (Przybilla \& Butler 2001), so the $\lambda$4471 and $\lambda$4922 helium lines are
calculated precisely but the diffuse $\lambda$4026 and $\lambda$4388 lines are calculated
less precisely. $\lambda$4713 is also reliable as it comes from a transition to level $n=4$.
As discussed in detail by Lyubimkov et al.\ (2004) regarding line broadening theory, the
$\lambda$4471 and $\lambda$4922 lines are to be preferred for helium abundance determinations.

There is growing evidence that, in the calculation of helium line profiles, \micro\ should be
carefully investigated (Gies \& Lambert 1992; McErlean, Lennon \& Dufton 1999; Lyubimkov,
Rostopchin \& Lambert 2004). Therefore, an extended grid of theoretical spectra were calculated
 for the \Teff s and \logg s found for the components of V453\,Cyg and for helium abundances
from $\epsilon_{\rm He} = 0.09$--$0.21$ in steps of $0.03$, and $\micro = 0$--20\kms\ in steps of 5\kms.

First, we have used the approach of Lyubimkov et al.\ (2004) to simultaneously determine \micro\ and
 $\epsilon_{\rm He}$. $\lambda$4471 and $\lambda$4922 were used for determining $\epsilon_{\rm He}$ 
as they have only a slight dependence on \micro. The $\lambda$4713 line has a strong dependence on
 \micro\ so was used to constrain this parameter once $\epsilon_{\rm He}$ was determined. As before,
 $\chi^2$ minimisation was used. The results are given in Table\,\ref{tab:helium} (columns 2 and 3).
 We find an enhanced helium abundance of $\epsilon_{\rm He} = 0.123 \pm 0.017$ for the primary star. 
It is evident that a large scatter between different \ion{He}{I} lines is present. The two largest 
values of the helium overabundance were obtained for the \ion{He}{I} $\lambda$4471 and $\lambda$4921 
lines, i.e.\ the lines Lyubimkov et al.\ (2004) favoured in determination of the helium abundance 
in their sample. Conversely, we find a normal abundance of $\epsilon_{\rm He} = 0.089 \pm 0.003$ 
for the secondary star. These results are discussed further in Section\,\ref{sec:discussion}.

\begin{table} \centering
\caption{\label{tab:helium} Helium abundances derived for the primary component
 of V453\,Cyg with the microturbulent velocity as free and fixed parameter. }
\begin{tabular}{lcccc} \hline
Line   & $\epsilon$(He) & $v_{turb}$ free  & $\epsilon$(He) &  $v_{turb}$ fixed    \\
\hline
4026  & 0.096$\pm$0.006  & 5$\pm$3  & 0.081$\pm$0.008 &  15 (fix)  \\
4388  & 0.113$\pm$0.008  & 7$\pm$2  & 0.075$\pm$0.010 &  15 (fix)  \\
4471  & 0.152$\pm$0.006  & 6$\pm$3  & 0.103$\pm$0.006 &  15 (fix)  \\
4713  & 0.118$\pm$0.009  & 7$\pm$2  &     -           &   -        \\
4921  & 0.134$\pm$0.009  & 8$\pm$2  & 0.086$\pm$0.012 &  15 (fix)  \\
\hline
mean  & 0.123$\pm$0.017  & 7$\pm$2 & 0.086$\pm$0.012 &  15 (fix)   \\
\hline \end{tabular} \end{table}

Secondly, we have determined the helium abundances using the value of \micro\ obtained from
metallic lines, in particular oxygen. In hot OB stars the \ion{O}{II} lines are relatively
strong and numerous, so are well suited to this purpose. \micro\ was derived by requiring
the abundances from individual \ion{O}{II} lines to be independent of the line strengths.
The equivalent widths were measured for 22 \ion{O}{II} lines which are not severely blended.
We find $\micro = 15 \pm 2$\kms.

It is well documented by previous studies that the \micro\ tends to be larger for lower
surface gravities (e.g.\ Kilian 1992; Vrancken et al.\ 2000; Daflon et al.\ 2004 and
references therein). In the sample of nine slowly-rotating $\beta$ Cephei stars recently
analyzed by Morel et al.\ (2008), three stars have $\micro > 10$\kms, whilst the full
range of \micro\ is 1--14\kms. All of these objects are in similar evolutionary state as
V453\,Cyg\,A ($\logg \sim 3.75$) so our derived \micro\ should be reliable.

Theoretical \ion{He}{I} line profiles were calculated for $\epsilon_{He} = 0.07$--0.15
and $\micro = 15$\kms. The \Teff\ and \logg\ were taken from Section\,\ref{sec:teff}.
The best fit by $\chi^2$ minimisation is a mean  helium abundance which is considerably
different from results obtained with \micro\ as a free parameter. The mean helium abundance
 in the primary's photosphere from the four \ion{He}{I} lines is $\epsilon(He) = 0.086
\pm 0.012$, almost equal to the canonical solar helium abundance [$\epsilon_{\odot}{\rm (He)}
= 0.089$]. The importance of this result is discussed in Section\,\ref{sec:discussion}.

\subsection{Metals}

\subsubsection{Primary star}

Determination of the \micro\ of the primary star also yielded an oxygen abundance of
$\log \epsilon(\rm O) = 8.42\pm0.087$. This estimate is obtained from 22 \ion{O}{II} lines,
and the quoted error is the rms around the mean. The abundances of the other elements were
estimated from fitting synthetic spectra to the disentangled spectrum of the primary star.
A grid of synthetic spectra were calculated for the parameters of the star and different
elemental abundances. The \micro\ was fixed to 15\kms, the value found from fitting the star's oxygen lines.

About ten lines were available for the estimates of the abundances for nitrogen and silicon.
The majority of N lines are blended by \ion{O}{II} lines, but show a good interagreement for
 an abundance $\log \epsilon({\rm N}) = 7.80 \pm 0.08$ (11 lines). By contrast, our measurement
of the Si abundance ($\log \epsilon({\rm Si}) = 7.27 \pm 0.14$, 9 lines) suffers from
a larger scatter in results for individual lines. The strong \ion{Si}{III} triplet at
$\lambda$4550 to $\lambda$4580 give larger abundances than the weak \ion{Si}{III} lines
on the far blue wing of H$\beta$. This could be due to an incorrect \micro\ or to contamination
by the H$\beta$ line. Interestingly, the analysis of $\beta$ Cephei stars by Morel et al.\ (2008),
also shows the largest dispersion in abundances for Si. This issue is also discussed by Trundle
et al.\ (2008) who used silicon lines for deriving \micro.

Only four carbon lines are present in our spectra: the strong \ion{C}{II} $\lambda$4267,
the weaker \ion{C}{III} $\lambda$4186, and two \ion{C}{II} lines at $\lambda$4650--$\lambda$4680
which are heavily blended with \ion{O}{II} lines. Our estimated C abundance (Table\,\ref{tab:abund4})
 suffers from the small number of useful lines and also from the quality of the available model
 atom (Trundle et al.\ 2008; Nieva \& Przybilla 2008).

The only visible magnesium line in the spectrum of the primary star is \ion{Mg}{II} $\lambda$4481, 
which is blended with \ion{Al}{III} $\lambda$4479. Aluminium is also detectable through other two 
lines (\ion{Al}{III} $\lambda$4512 and $\lambda$4532). The abundances of Mg and Al were derived by
 calculating theoretical spectra fitting blended line profiles in the 4477--4484\,\AA\ region. 
The best match was found for $\log \epsilon({\rm Mg}) = 7.58$ and $\log \epsilon({\rm Al}) = 6.10$.
 The other two \ion{Al}{III} lines give abundances consistent with this.

Beside intrinsic (random) errors, several systematic errors affect the determination of abundances,
such as uncertainties in the atmospheric parameters (\Teff, \logg\ and \micro) and in fundamental
 atomic data ($gf$ values). Morel et al.\ (2006) and Trundle et al.\ (2007) have discussed
 these problems, and we have adopted their approach in estimating the overall uncertainties in
 our abundance measurements. We repeated our abundance analysis using synthetic spectra with
 atmospheric parameters modified by $\Delta T = 1000$K, and $\Delta\micro = 3$\kms. The uncertainty
in the \logg s of V453\,Cyg is only 0.01\,dex, so is negligible here. The overall
uncertainties in the abundances for the primary star (Table\,\ref{tab:abund4}) include
the line-to-line scatter and the effects of changes in \Teff\ and \micro. The error budget was
 calculated only for O, N and Si, for which sufficient lines are present in the primary star's
spectrum. For the other elements, a conservative upper limit on the total error of 0.20 dex
is assigned.

\subsubsection{Secondary star}

We have so far concentrated on the primary star, as its mass and evolutionary state are astrophysically
 very interesting. We now measure elemental abundances for the secondary star by differential analysis. 
In the study of the dEB V578\,Mon, PH05 devised a complete methodology for the determination of 
elemental abundances from disentangled spectra of high-mass stars. Special attention was given to 
the minimisation of systematic errors arising from continuum placement. The abundance determinations
 were made relative to rotationally broadened spectra of a slowly-rotating star which was otherwise 
similar to the components of the binary. A detailed examination of the spectra and comparison to 
template spectra made it possible to extract the blended line profiles. Here we use the same 
methodology and wavelength ranges as in PH05.

V453\,Cyg\,B ($\Teff = 26\,200 \pm 500$\,K) closely resembles V578\,Mon\,B ($\Teff = 26\,400 \pm 
400$\,K). Their surface gravities differ by only 0.15\,dex and their rotational velocities are similar, 
so a direct comparison of their spectra is possible. Fig.\,\ref{fig:v578v453} is a plot of the 
$\lambda$4635--$\lambda$4657 spectral region for the two stars, which contains a complex blend 
of \ion{O}{II}, \ion{N}{II}, \ion{N}{III}, \ion{C}{II}, \ion{C}{III} and \ion{Si}{IV} lines.

There is striking similarity of both spectra throughout the observed spectral windows. Some slight 
differences can be found, and it is almost certain that these are intrinsic to the stars rather than
 an artifact of {\sc spd}. For  comparison synthetic spectra have been calculated for elemental 
abundances of an 'average' main-sequence B star; mean values of the abundances are $\log \epsilon(\rm C) 
= 8.27$, $\log \epsilon(\rm N) = 7.62$, $\log \epsilon(\rm O) = 8.57$, $\log \epsilon(\rm Mg) = 7.48$ 
and $\log \epsilon(\rm Si) = 7.25$ (Daflon, Cunha \& Becker 2004). The spectra of both stars closely 
follow the characteristics of the synthetic spectrum. The elemental abundances of V578\,Mon\,B resemble
 'average' B-dwarf abundances in the quoted errors (PH05, their table\,3).

As V453\,Cyg is a probable member of the open cluster NGC\,6871, which is part of the Cyg\,OB3 
association, we can compare our results with the work of Daflon et al.\ (2001). These authors determined 
elemental abundances of OB stars in five OB associations. Spectra of eights objects from Cyg\,OB3 were 
obtained, but final analyses were performed for only three sharp lined targets\footnote{V453\,Cyg was 
included in the study by Daflon et al.\ (2001), who found $\Teff = 29\,900$\,K and $\logg = 4.53$. 
The composite nature of the spectra was not accounted for, so these quantities are not reliable.}.

The results of Daflon et al.\ (2001) for the stars in Cyg\,OB3 are:
$\log \epsilon(\rm C) = 8.15\pm0.23$,
$\log \epsilon(\rm N) = 7.68\pm0.19$,
$\log \epsilon(\rm O) = 8.54\pm0.14$,
$\log \epsilon(\rm Mg) = 7.63\pm0.24$,
$\log \epsilon(\rm Si) = 7.22\pm0.11$.
Our quantities for both V453\,Cyg\,A and V453\,Cyg\,B are in excellent agreement with these figures.

\begin{table} \centering
\caption{\label{tab:abund4} Abundances derived for the components of V453\,Cyg}
\begin{tabular}{lccc} \hline
Abundance  &  Primary  & Secondary  & OB stars  \\
\hline
H/He                      & 0.094$\pm$0.015\,(4)     & 0.098$\pm$0.011    &   0.10  \\
$\log \epsilon({\rm C})$  &   8.15$\pm$0.20\,(2)     &  8.30$\pm$0.12     &   8.2   \\
$\log \epsilon({\rm N})$  &   7.80$\pm$0.10\,(11)    &  7.60$\pm$0.10     &   7.6   \\
$\log \epsilon({\rm O})$  &   8.58$\pm$0.12\,(22)    &  8.55$\pm$0.05     &   8.5   \\
$\log \epsilon({\rm Mg})$ &   7.58$\pm$0.20\,(1)     &  7.50$\pm$0.10     &   7.4   \\
$\log \epsilon({\rm Si})$ &   7.27$\pm$0.16\,(9)     &  7.25$\pm$0.12     &   7.2   \\
$\log \epsilon({\rm Al})$ &   6.12$\pm$0.20\,(3)     &   -                &   6.1   \\
\hline \end{tabular} \end{table}


\section{Discussion}                    \label{sec:discussion}

Due to the serious problems of line blending and high rotational velocities, abundances have been measured
 for only a few high-mass stars in double-lined binary systems. Prior to the invention of {\sc spd} 
(Simon \& Sturm 1994) some helium abundances in high-mass binaries were determined from equivalent 
widths measured using Gaussian fitting (Lyubimkov 1998; Pavlovski 2004 and references therein). 
The use of {\sc spd} has allowed helium abundances to be measured for four close binaries: DH\,Cep 
(Sturm \& Simon 1994), Y\,Cyg (Simon et al.\ 1994), V578\,Mon (Hensberge et al.\ 2000), and DW\,Car 
(Southworth \& Clausen 2007). Preliminary results have also been announced for V380\,Cyg (Pavlovski 
et al.\ 2005), and V478\,Cyg and CW\,Cep (Pavlovski \& Tamajo 2007). Finally, Freyhammer et al.\ (2001)
 have derived helium abundances for the components of CPD\,$-$59$^\circ$2628 by fitting theoretical spectra.

Lyubimkov et al.\ (2004) found an enrichment of helium during the MS evolution of their sample of stars,
 although many theoretical evolutionary models (e.g.\ Schaller et al.\ 1992) predict no surface enrichment
 of helium or other CNO-cycle elements. However, more recent theoretical studies (Maeder \& Meynet 2000; 
Heger \& Langer 2000) have found that rotationally induced mixing can affect the chemical composition of 
the surface layer of high-mass stars even whilst they are on the MS. Due to CNO processing in the stellar 
core, He and N should be enriched and C and O depleted. The strength of these effects depends on initial 
rotational velocity, so observational constraints can be used to both calibrate theoretical models and 
check their predictions (e.g.\ Venn et al.\ 2000).

The observations of 100 Galactic B stars by Lyubimkov et al.\ (2000, 2004) have confirmed that helium 
overabundance is correlated with stellar age expressed as a fraction of its MS lifetime. Large 
spectroscopic surveys of OB stars in our Galaxy and the Large and Small Magellanic Clouds (Hunter 
et al.\ 2008) have further confirmed this effect at a range of metallicities. Unfortunately, 
investigations using single stars suffer from the low accuracy with which their masses, radii 
and ages can be measured. Further work must therefore concentrate on binary stars, where all these
 properties can be measured to within a few percent (e.g.\ SMS04).

The subsample of high-mass stars in the sample of OB stars studied by Lyubimkov et al.\ (2004) show 
helium enrichment during their MS lifetimes, which has been found previously for close binaries 
(see Lyubimkov 1998 and references therein). In two recent studies in which large samples of B stars 
were analysed, progressively increasing helium enrichment from the ZAMS to the TAMS was found 
(Lyubimkov et al.\ 2004, Huang \& Gies 2006). Both studies saw this effect over their whole mass 
ranges, with large enrichments for high-mass stars. Moreover, Huang \& Gies noted that the effect 
is large among the faster rotators (see their fig.\ 12). Lyubimkov et al.\ derived \micro\ both from 
helium lines and from \ion{O}{II}, while Huang \& Gies (working with only a limited spectral range 
around H$\gamma$) accepted $\micro\ = 2$\kms\ for all their sample after considering the discussion 
of \micro\ in Lyubimkov et al.\ (2004). Therefore, the agreement achieved by these studies may depend 
on their use of a common \micro.

So far, in the sample of OB binaries studied by {\sc spd} and cited above, no MS star has been found 
to have a helium enrichment. To this list we add now both components of V453\,Cyg. The ages of the 
stars on the list lie between 0 and 0.8 of their MS lifetimes, with V453\,Cyg\,A contributing the 
upper limit. Possible explanations for the non-detection of any changes in elemental abundances are 
moderate rotational velocity and tidal effects between the components. In this context it is 
interesting that detailed abundance analyses of about a dozen $\beta$ Cephei stars (Morel et al.\ 
2006) has given an average helium abundance of $\epsilon({\rm He}) = 0.087 \pm 0.012$, with a range 
from $0.075$ to $0.100$. These objects are slowly-rotating stars in advanced phases of their MS 
lifetimes and do not show enhanced helium abundances. The \Teff\ and \logg\ of V453\,Cyg\,A fall 
within the range of their sample, and this star also does not show enhanced helium. These two 
results are in excellent agreement and corroborate previous finding of Herrero et al.\ (1992), 
who found a very clear dependence of helium enrichment on rotational velocity, but not on fractional 
lifetime on the MS, for O-type stars.

In Section\,\ref{sec:he} we determined the He abundance in V453\,Cyg\,A by two approaches: (i) 
simultaneous line-profile fitting allowing \micro\ to be a free parameter; (ii) fixing \micro\ to 
the value obtained from \ion{O}{II} lines. When using a fixed \micro\ we found a normal helium 
abundance ($\epsilon({\rm He)} = 0.086 \pm 0.012$), but when \micro\ was a free parameter we found 
a helium overabundance ($\epsilon({\rm He}) = 0.123 \pm 0.017$) and also erratic metal abundances. 
We therefore reject the helium enrichment scenario and caution that it is crucial to carefully 
consider the value of \micro\ in similar studies.


\section{Summary}

Binary and multiple stellar systems are our primary source of fundamental measurements of the basic 
physical properties of stars. Modern observational data and analytical tools allow us to determine 
stellar masses and radii with accuracies of 1\% or better, which in turn yields surface gravity 
measurements accurate to a few hundredths of a dex. When combined with advanced spectroscopic and 
abundance analysis techniques it is now possible to perform detailed tests of the rotation and 
magnetic effects included in modern theoretical evolutionary models.

In this work we have studied the eclipsing double-lined spectroscopic binary V453\,Cyg, a system 
which contains two stars with accurately measured masses and radii. V453\,Cyg\,A is approaching the 
TAMS and V453\,Cyg\,B has completed about half of its MS phase. These characteristics make V453\,Cyg 
ideal for testing the MS chemical evolution of high-mass stars, which is affected by convection, 
rotational mixing, and magnetic fields.

Through {\sc spd} of several sets of phase-resolved high-resolution spectroscopy we have obtained 
the individual spectra of the two stars. By fitting the hydrogen line profiles with non-LTE synthetic
 spectra, we have measured the effective temperatures for the two stars to be ${\Teff}_{\rm A} = 
27\,800 \pm 400$\,K and ${\Teff}_{\rm B} = 26\,200 \pm 500$\,K. The high precision of these 
measurements was achieved because the surface gravities of the two stars are known to within 
0.01\,dex (SMS04). The \Teff s are larger than have been found previously, and are supported by 
the Str\"omgren photometric indices of the system.

The new \Teff\ values imply a distance of $1680 \pm 30$\,pc to V453\,Cyg (using theoretical 
bolometric corrections), which is in good agreement with most literature values but still shorter 
than the distance of $2140 \pm 70$\,pc found for NGC\,6871 by Massey et al.\ (1995). The predictions 
of theoretical stellar evolutionary models are unable to match the properties of V453\,Cyg.

Using the disentangled spectra we have measured the helium and metal abundances in the photospheres 
of the two stars by fitting non-LTE synthetic spectra to the observed line profiles. The metal abundances
 for the secondary star were estimated by differential analysis to a template star of similar \Teff\ 
and \logg. We find that the helium in the photosphere of the primary has a solar abundance if we adopt 
a microturbulent velocity obtained from metallic lines. This result does not contradict evolutionary
 calculations (Meynet \& Maeder 2000; Heger \& Langer 2000) because a slight helium enrichment is 
predicted only for highly rotating stars, and/or after TAMS. Also, as is expected for the stars which
 were formed from the same interstellar material, no significant difference in the chemical composition
 between the two components is apparent. Their photospheric abundances resemble those of the typical
 Galactic OB stars (c.f.\ Herrero 2003 and references therein).

We are planning to extend our analysis to larger numbers of spectral lines, which will require new
 spectroscopic observations with a wider wavelength coverage. This will make it possible to refine
 the present abundance analysis, and to address more thoroughly the question of which \micro\ to 
adopt for different ions. \'Echelle spectroscopic observations will be well suited to this task.


\section*{Acknowledgements}

We would like to thank Dr.\ Klaus Simon who kindly put spectra of V453\,Cyg, obtained at
German-Spanish Astronomical Centre on Calar Alto, Spain, at our disposal. We would also like
to thank Dr.\ Herman Hensberge for contributing a spectrum of NGC\,2244 \#201. We are
grateful to Prof.\ Artemio Herrero for careful and critical reading of the first draft.
Thanks are also due to Dr.\ Simon Daflon and Dr.\ Thierry Morel for illuminating discussions.
Research by KP is funded through a research grant from Croatian Ministry of Science \& Education.
JS acknowledges financial support from STFC in the form of a postdoctoral research associate
position.


\label{lastpage}

\end{document}